\documentclass[useAMS,usenatbib]{mn2e}

\usepackage{graphicx}
\graphicspath{{converted_graphics/}}
\begin{document}
\title[Supermassive Stars]{Evolution of supermassive stars as a pathway to black hole formation}

\author[M.~C.~Begelman] {Mitchell C. Begelman$^{1,2}$  \\
$^1$JILA, University of Colorado at Boulder, 440 UCB, Boulder, CO 80309-0440 \\
$^2$Department of Astrophysical and Planetary Sciences, University of Colorado \\ 
\tt e-mail: mitch@jila.colorado.edu}

\maketitle
  
\begin{abstract}
Supermassive stars, with masses $\ga 10^6 M_\odot$, are possible progenitors of supermassive black holes in galactic nuclei.  Because of their short nuclear burning timescales, such objects can be formed only when matter is able to accumulate at a rate exceeding $\sim 1 M_\odot$ yr$^{-1}$.  Here we revisit the structure and evolution of rotationally-stabilized supermassive stars, taking into account their continuous accumulation of mass and their thermal relaxation.  We show that the outer layers of supermassive stars are not thermally relaxed during much of the star's main sequence lifetime. As a result, they do not resemble $n=3$ polytropes, as assumed in previous literature, but rather consist of convective (polytropic) cores surrounded by convectively stable envelopes that contain most of the mass.  We compute the structures of these envelopes, in which the equation of state obeys $P/\rho^{4/3} \propto M^{2/3}(R)$, where $M(R)$ is the mass enclosed within radius $R$.  By matching the envelope solutions to convective cores, we calculate the core mass as a function of time.  We estimate the initial black hole masses formed as a result of core-collapse , and their subsequent growth via accretion from the bloated envelopes (``quasistars") that result.  The seed black holes formed in this way could have typical masses in the range $\sim 10^4-10^5 M_\odot$, considerably larger than the remnants thought to be left by the demise of Pop III stars.  Supermassive black holes therefore could have been seeded during an epoch of rapid infall considerably later than the era of Pop III star formation.
\end{abstract}

 
\section{Introduction}

Supermassive stars were proposed by Hoyle \& Fowler (1963a,b) as a means to meet the prodigious energy requirements of radio galaxies (Burbidge 1958), and slightly later as a model for quasars.  Hoyle \& Fowler recognized that such objects could not persist for much more than a million years, if sustained by hydrogen burning, and suggested that they might collapse to form black holes once their nuclear fuel was exhausted.  The pulsational stability of supermassive stars is a crucial issue, because they are radiation pressure-dominated and therefore have an adiabatic index very close to $4/3$ (which yields neutral stability to radial pulsations for a Newtonian, self-gravitating body with no rotation).  Small general relativistic corrections have a destabilizing effect, preventing nonrotating stars more massive than a few $\times 10^5 \ M_\odot$ from attaining a phase of stable hydrogen burning before collapsing (Iben 1963; Fowler 1964).  However, a dynamically insignificant level of rotation --- especially differential rotation --- can stabilize stars as massive as $\sim 10^8 M_\odot$ or more (Fowler 1966).  Additional stabilizing effects due to magnetic fields and turbulence were later considered by other authors (e.g., Bisnovatyi-Kogan, Zel'dovich \& Novikov 1967;   Ozernoy \& Usov 1971). 

In this paper, we are interested in whether stable supermassive stars might be precursors of seed black holes that eventually grow to large enough mass to power quasars and populate the nuclei of present-day massive galaxies.  As we will argue in \S~5, the holes created inside a supermassive star are likely to have masses of a few percent of the star's final mass.  Thus, to obtain seed black holes with masses $\sim 10^4-10^5 M_\odot$, comfortably larger than the seeds probably left behind by the collapse of Population III stars, we need to consider supermassive stars with masses $\ga 10^6 M_\odot$.  We will focus on this mass range in our analysis.\footnote{Note, however, that the term {\it supermassive star} is often used to refer to the mass range $M_*\ga 5\times 10^4  M_\odot$ (Fuller, Woosley \& Weaver 1986).} 

Once the existence of a stable supermassive star is postulated, its structure would appear to be quite simple. In order for radiation pressure to support the star against gravity, the radiative flux seen by every element of mass in the star must equal the {\it local} Eddington limit, $F_E =  GM(R)c/\kappa r^2$, where $\kappa$ is the local opacity and $M(R)$ is the mass enclosed within $R$.  The opacity inside a supermassive star is likely to be roughly uniform, with electron scattering dominant everywhere, but $M(R)$ is a monotonic function of radius.  Since the entire luminosity is produced by thermonuclear reactions within a small region of the core, the star must be highly convective with the fraction of the total luminosity carried by convection varying as $L_{\rm conv}/L_{\rm tot} = 1 - M(R)/M_*$, where $M_*$ is the total mass of the star.  If the convection is efficient, then a supermassive star should be described accurately by an $n=3$ ($\gamma = 4/3$) polytrope, according to the Lane-Emden equation.  

Such models, however, beg the question of how a supermassive star might realistically form.  Creating a $10^6 M_\odot$ star requires the very rapid accumulation of gas. Since the thermonuclear timescale for a star burning hydrogen at the Eddington limit is $\sim2$ Myr, independent of mass, and will be less if only part of the fuel is burned, the infall rate required to create a supermassive star of mass $M_*$ must exceed $0.5 (M_*/10^6 M_\odot) \ M_\odot$ yr$^{-1}$.  This is orders of magnitude larger than the rate at which matter condenses to form normal stars within molecular clouds, or even the rate ($\sim 10^{-3} M_\odot$ yr$^{-1}$) at which matter came together to form Pop III stars in pregalactic dark matter halos.   

The maximum rate at which matter can collect is given by $\sim v^3/G = 0.2 (v/10 \ {\rm km \ s}^{-1})^3 M_\odot$ yr$^{-1}$, under the assumption that the gas is self-gravitating and collapses at its free-fall speed $v$.  In effect, this means that supermassive stars can only form in systems with virial temperatures exceeding $10^4$ K.  There is considerable controversy over whether such high rates of inflow can occur without most of the gas fragmenting and forming stars before reaching the center.  Despite early suggestions that fragmentation is avoidable only if the gas temperature remains close to the virial temperature (Bromm \& Loeb 2003; Begelman, Volonteri \& Rees 2006) --- and thus only in systems lacking both metals and molecular hydrogen ---  recent simulations suggest that fragmentation may be suppressed even when the gas is much colder (Wise, Turk \& Abel 2008; Levine et al. 2008; Regan \& Haehnelt 2009), possibly due to the continuous generation of supersonic turbulence (Begelman \& Shlosman 2009).
 
Conditions of rapid infall are ideal for producing the high levels of entropy required inside supermassive stars.  Indeed, simple estimates show that the initial entropy of the gas joining the star can be much larger than the value required for equilibrium.  We argue that the nature of radiation-pressure support provides a mechanism for automatically regulating the entropy of the gas joining the supermassive star, so that it is neither too large nor too small. This is due to the existence of a ``trapping radius" within the infalling gas, outside of which radiative diffusion can release excess entropy.  Gas with too much entropy will be forced to expand until it gives up the excess. However, this does {\it not} imply that supermassive stars should have uniform specific entropy and therefore resemble polytropes.

In this paper we study the evolution and fate of supermassive under the assumption that they grow by rapid, continuous infall. In \S\S~2 and 3, we argue that gas joins the star with increasing specific entropy as a function of time.  Since the Kelvin-Helmholtz timescale is longer than the age of the star for early times and high infall rates, much of this this entropy stratification is preserved over the life of the star.  We therefore conclude that supermassive stars are not necessarily well-represented by $n=3$ polytropes, but rather can have a more complex structure with a convective (polytropic) core surrounded by a convectively stable envelope that contains most of the mass.  The entropy in the envelope roughly satisfies $P/\rho^{4/3} \propto M^{2/3}(r)$ (we term this entropy law ``hylotropic").  Hydrogen burning in the core starts when the star's mass and entropy are both relatively low, and adjusts to sustain the star through its more massive stages (\S~4).  

In \S~5, we discuss the formation of the seed black hole after the exhaustion of core hydrogen, and its subsequent growth inside the remnant of the supermassive star.  Because of the likely importance of rotation in the collapsing core, only a small fraction of the supermassive star collapses to a black hole initially.  The energy liberated during the formation of the black hole inflates the remainder of the star into a bloated object that resembles a red giant, which we have previously termed a ``quasistar" (Begelman et al.~2006; Begelman, Rossi \& Armitage 2008).  The black hole continues to grow by accretion from the quasistar envelope, until the photospheric temperature drops to the point where the quasistar undergoes an ``opacity crisis" and disperses under the influence of radiation pressure.  We summarize our results in \S~6 and comment on the possible contribution of supermassive stars to the ionizing radiation field at high redshifts.

\section{Self-consistency of Fully Convective Models}

A thermally relaxed supermassive star must be fully convective and therefore well-represented by an $n=3$ ($\gamma = 4/3$) polytrope (Hoyle \& Fowler 1963a).  The structure is determined by the solution, $\theta_3 (x)$, of the Lane-Emden equation:
\begin{equation}
\label{LaneEmden}
{1\over x^2}{d\over dx}\left(x^2{d\theta_3\over dx}\right) = -\theta_3^3
\end{equation}
with boundary conditions $\theta_3(0) = 1$, $\theta_3'(0)=0$, where $x \equiv R/R_3$ is the radius normalized to $R_3 = (P_c/\pi G \rho_c^2)^{1/2}$. $P_c$ and $\rho_c$ are the central pressure and density, respectively, with the pressure and density elsewhere being given by $P(x) = P_c \theta_3^4 (x)$ and $\rho(x) = \rho_c \theta_3^3 (x)$.  Density is related to pressure via an entropy parameter $K\equiv P/\rho^{4/3}$, which depends uniquely on the mass of the star: $K =  0.364 GM_*^{2/3}$. The dimensionless radius of the star is given by $x_* = 6.897$, corresponding to
\begin{equation}
\label{polyradius}
R_* = 5.8 \times 10^{13} m_{*,6}^{1/2} T_{c,8}^{-1} \ {\rm cm},
\end{equation} 
where $m_{*,6}= M_*/10^6 M_\odot$ and $T_c = 10^8 T_{c,8}$ K is the central temperature.  Rather than assuming that the star simply exists with a fixed mass, we assume that the star's mass is growing at a rate $\dot M_* = \dot m_* \ M_\odot$ yr$^{-1}$, so that $m_{*,6}(t) = \dot m_* \ t_{\rm Myr}$.

Given the growth rate of the star, we can assess whether the star is thermally relaxed by comparing $R_*$  to the ``trapping radius," $R_{\rm tr}$, which separates the outer region where the infalling gas leaks radiation, from the inner region where the flow is approximately adiabatic (Begelman 1978). For electron scattering opacity with $\kappa = 0.34$ cm$^2/$g,  we have 
\begin{equation}
\label{rtrap}
R_{\rm tr} = {\kappa \dot M_* \over 4\pi c} = 5.7 \times 10^{13} \dot m_* \ {\rm cm}.
\end{equation} 
Suppose $R_*$ lies inside the trapping radius.  In this case, the infalling gas must already be adiabatic by the time it joins the star.  This is problematic, because the specific entropy of the gas in an $n=3$ polytrope (as measured, e.g., by $K$) is uniquely determined by the mass.  Matter about to join the star will generally have much higher entropy than the required value. (This can be ascertained, in a specific case, by comparing the ram pressure with the density of gas in free fall --- but it would be true in most circumstances.) Therefore, infalling matter {\it must} lose entropy in order to join a polytropic star.  But if $R_* < R_{\rm tr}$, the gas behaves adiabatically near the star and such entropy loss is impossible.  In this case the star cannot be thermally relaxed.  On the other hand, thermal relaxation is possible if $R_* > R_{\rm tr}$.

Comparing equations (\ref{polyradius}) and (\ref{rtrap}), we see that supermassive stars of a given mass can  become thermally relaxed only after a certain amount of time has passed,
\begin{equation}
\label{rtrap2}
t_{\rm Myr} >  m_{*,6}^{1/2} T_{c,8}, 
\end{equation}
or equivalently, for a given accretion rate, if $t_{\rm Myr} >  \dot m_* T_{c,8}^2$.  The ``main sequence" lifetime for a fully convective star is $\approx 2$ Myr, where we have assumed a hydrogen mass fraction $X = 0.75$.  This implies that no star more massive than $\sim 4 \times 10^6 T_{c,8}^{-2} M_\odot$ can be thermally relaxed during its lifetime. 

\section{Partially Convective Supermassive Stars}

If matter falling onto a growing supermassive star has the ``wrong" entropy to maintain a polytrope, what kind of structure results?   It appears that the acquisition of entropy can be a self-regulating process.  If gas with too much entropy per unit mass is added to an $n=3$ polytrope, for example, the star will expand to the point where its radius exceeds the trapping radius and it can radiate away the excess entropy.  If, conversely, gas with too little entropy is added to the polytrope, the star will shrink until thermonuclear reactions increase the entropy in the interior.  Since a supermassive star is built up by adding gas with ``too much" entropy, this implies that the radius of a supermassive star is of order the trapping radius associated with the infall rate. For constant $\dot M$, the radius of a supermassive star is independent of mass.  This is impossible if the entropy is uniform throughout the star (that would be a polytrope, with $R_* \propto M_*^{1/2}$), but it is possible if the specific entropy increases with radius. 

At each stage in the star's growth, the entropy of the newly-added gas adjusts to the ``correct" value for the star's current mass.  As more matter is added, the earlier layers are squeezed to smaller radii, and their entropies are frozen in.  From the homology scalings $P \sim GM^2/R^4$ and $M\sim \rho R^3$, which must be satisfied for each layer of mass, we see that the entropy must be an increasing function of enclosed mass, $P/\rho^{4/3} \propto M^{2/3}$. Since mass increases with radius, this implies that the envelopes of unrelaxed stars are stable to convection.  

We now construct quantitative models for these unrelaxed envelopes, and show how they can be matched to convective cores. Because the equation of state depends explicitly on the enclosed mass, it is convenient to use Lagrangian variables.  Assuming hydrostatic equilibrium, the equations are:
\begin{equation}
\label{mass}
4\pi \rho R^2 R' = 1   \hfill {\rm (mass\ conservation)} 
\end{equation}
\begin{equation}
\label{momentum}
P' = - {GM \over 4\pi R^4 }    \hfill {\rm (hydrostatic\ equilibrium)}, 
\end{equation}
where a prime denotes differentiation with respect to $M$, to which we add the equation of state
\begin{equation}
\label{eossms}
P = A\rho^{4/3} M^{2/3}, 
\end{equation}
where $A$ is a constant.

\subsection{Hylotropes}

Equation (\ref{eossms}) is a special case of a more general class of equations of state of the form 
\begin{equation}
\label{eoshyl}
P = A\rho^{4/3} M^\alpha. 
\end{equation}
Because of the dependence on mass, we propose to call such equations of state {\it hylotropic} and the derived structures {\it hylotropes} (from Gk.~{\it hyle}, ``matter" $+$ {\it tropos}, ``turn").\footnote{We thank A.~Accardi and G.~Lodato for suggesting this compound word and providing its etymology.}  
By eliminating $\rho$ through equations (\ref{mass}), (\ref{momentum}), and (\ref{eoshyl}), we obtain
\begin{equation}
\label{hyl1}
{P'\over P} = {\alpha\over M}- {8\over 3}{R'\over R} - {4\over 3}{R''\over R'} = - {(4\pi)^{1/3}GM^{1-\alpha}\over A}  \left({R'\over R}\right)^{4/3} .  
\end{equation}  
Eliminating $R$ in favor of $w \equiv M R'/R$, we obtain
\begin{equation}
\label{w1}
M {w'\over w} = 1 + {3\alpha\over 4}-3w + \eta M^{{2\over 3} - \alpha}w^{4/3} , 
\end{equation}
where $\eta \equiv 3(4\pi)^{1/3} G/4 A$. Equation (\ref{w1}) reduces to a Lagrangian equivalent of the Lane-Emden equation for $n=3$ polytropes when $\alpha = 0$.  For the case of interest here, $\alpha = 2/3$, we have 
\begin{equation}
\label{w2}
M {w'\over w} = {3\over 2}(1-2w) + \eta w^{4/3} , 
\end{equation}
which is unique in having no preferred mass scale ($\eta$ is dimensionless). Equation (\ref{w2}) admits power-law solutions, $R\propto M^w$  with $w =$ const., provided that the right-hand-side has a real root.  This occurs for $\eta < \eta_{\rm crit} = 9/2^{7/3}\approx 1.7858$, i.e., for sufficiently large entropy (since $\eta$ scales inversely with $A$).  The power-law solutions are unbounded and therefore unphysical, just as $n=3$ polytropes are if the entropy is too large for the mass, i.e., if $K >  0.364 GM_*^{2/3}$.  

From eq.~(\ref{mass}), we see that 
\begin{equation}
\label{wdef}
w = {1\over 3}{\bar \rho \over \rho} \ , 
\end{equation}
where $\bar \rho$ is the mean density inside $R$.  Since $\rho$ must be a monotonically decreasing function of radius, we must have $w \geq 1/3$ everywhere.  All solutions satisfying this constraint at some inner boundary $M_m$ tend to power-laws at  $M \gg M_m$ when $\eta < \eta_{\rm crit}$; therefore, we are interested only in models with $\eta > \eta_{\rm crit}$, all of which are bounded.   

The bounded solutions all have $w$ increasing monotonically with $M$ and $R$; the outer edge of the envelope is reached where $w$ diverges. In practice, it is easiest to treat $w$ as the independent variable, computing $M(w)$ and $R(w)$ by integrating 
\begin{equation}
\label{w3}
{d \ln M \over d w} = w^{-1}\left[{3\over 2}(1-2w) + \eta w^{4/3}\right]^{-1} , 
\end{equation}
\begin{equation}
\label{w4}
{d \ln R \over d w} = \left[{3\over 2}(1-2w) + \eta w^{4/3}\right]^{-1} . 
\end{equation}
The normalizations of $M$ and $R$ are arbitrary; one need only specify an initial value of $w$ and integrate outward.
 
\subsection{Matching to a Convective Core}

Suppose that an $\alpha = 2/3$ hylotropic envelope matches to a convective core at $M_m$, $R_m$. The equation of hydrostatic equilibrium implies that $P$ and $P'$ are continuous, while the continuity of the specific entropy implies that $\rho$ is continuous as well.  The mass conservation equation then implies that $w = MR'/R$ is continuous. Only $\rho'$ changes discontinuously across the boundary. Matching the hylotropic and polytropic equations of state at the boundary gives a relation between the polytropic entropy parameter $K$ and the hylotropic entropy parameter $\eta$:
\begin{equation}
\label{Keta}
K = A M_m^{2/3} = {3 (4\pi)^{1/3} GM_m^{2/3} \over 4 \eta} .
\end{equation}   

The mass enclosed within the dimensionless matching radius $x_m$ is 
\begin{equation}
\label{enclmass}
M_m = {4 K^{3/2}\over \pi^{1/2} G^{3/2}}\int^{x_m}_0  \theta_3^3 x^2 dx .
\end{equation}
Substituting for $K$ from eq.~(\ref{Keta}), we find that $M_m$ cancels out and we are left with a relationship between $x_m$ and $\eta$:
\begin{equation}
\label{etaxm}
\eta = 3 \left[\int^{x_m}_0  \theta_3^3 x^2 dx \right]^{2/3}.
\end{equation}
From eq.~(\ref{wdef}), we see that the matching radius also determines the value of $w$ at the match point,
\begin{equation}
\label{wxm}
w_m =  {1\over x_m^3 \theta_3^3(x_m)} \int^{x_m}_0  \theta_3^3 x^2 dx .
\end{equation}

Thus, by selecting a matching radius $x_m$ we specify both the entropy scaling parameter of the envelope (i.e., $\eta$) and the initial value of $w_m$ to use in integrating equations (\ref{w3}) and (\ref{w4}).  The critical entropy parameter $\eta_{\rm crit}$ corresponds to a critical matching radius $1.2957 < x_{\rm crit} < 1.2958$ and a critical value of $w_m = w_{\rm crit} \approx 0.455$.   There are no bounded envelope solutions for $x_m < x_{\rm crit}$ ($w_m <w_{\rm crit}$), since these correspond to $\eta < \eta_{\rm crit}$.

As the critical values of these parameters are approached from above, the ratios of the envelope mass to the core mass ($M_*/M_m$) and the envelope radius to the core radius ($R_*/R_m$) diverge.  For example, for $x_m = 1.2958$, numerical integration of equations (\ref{w3}) and (\ref{w4}) give $M_*/M_m = 2.1 \times 10^8$ and $R_*/R_m = 3.6 \times 10^{16}$, respectively. However, these ratios decline extremely rapidly as the parameters increase only slightly above their critical values. Fig.~1 shows that all interesting ratios of envelope-to-core mass and radius are contained within a narrow range of matching radii and entropy parameters.  

\begin{figure}
  \centering
  \includegraphics[bb=0 0 336 243,width=3.2in,height=3.2in,keepaspectratio]{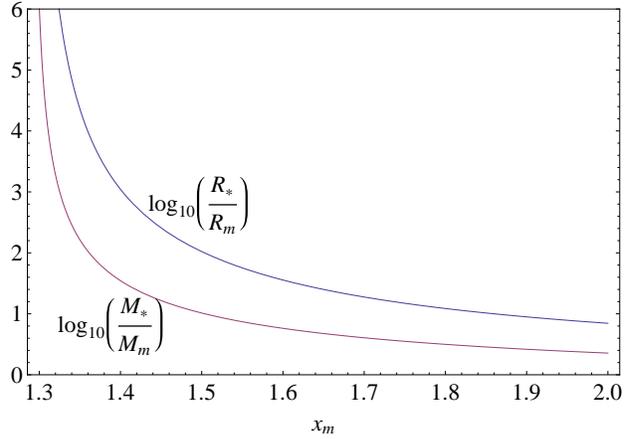}
  \caption{Ratios of envelope-to-core mass (lower curve) and radius (upper curve) for hylotropic envelopes matched to polytropic ($n=3$) cores, as a function of the dimensionless matching radius $x_m$.  As $x_m$ increases from 1.3 to 2, the mass ratio decreases from $10^6$ to just 2.27. Over the same interval, the entropy parameter $\eta$ goes from 1.80 to 3.09.}
  \label{fig:SMSEnvelope1}
\end{figure}

Envelope profiles are easily computed using equations (\ref{mass}) and (\ref{eossms}).  Figures 2 and 3 show plots of the normalized density and pressure, 
\begin{eqnarray}
\label{normdensity}
{\rho\over \rho_m}& = &{M\over M_m}\left({R\over R_m}\right)^{-3}{w_m\over w}\ ; \nonumber \\ 
 &  &  \nonumber \\
{P\over P_m} & = & \left({M\over M_m}\right)^2\left({R\over R_m}\right)^{-4}\left({w_m\over w}\right)^{4/3},
\end{eqnarray}
for $M_*/M_m = 10, 100, 10^3$. The density decreases with radius roughly as $R^{-5/2}$ while the pressure decreases as $R^{-3}$.  However, a power-law is a poor approximation: the slopes vary continuously.  This curvature on the log--log plot is more evident if one plots density or pressure vs.~mass.

\begin{figure}
  \centering
  \includegraphics[bb=0 0 377 256,width=3.23in,height=3.23in,keepaspectratio]{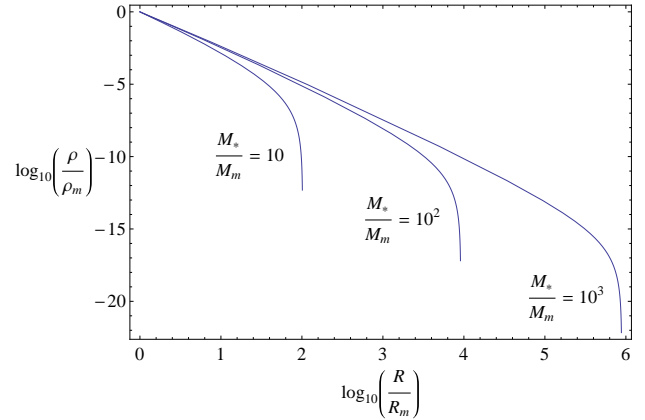}
  \caption{Density vs.~radius for hylotropic envelopes surrounding convective cores, normalized to conditions at the matching radius, for three ratios of envelope mass to core mass.}
  \label{fig:SMSEnvelope3}
\end{figure}
\begin{figure}
  \centering
  \includegraphics[bb=0 0 388 263,width=3.23in,height=3.23in,keepaspectratio]{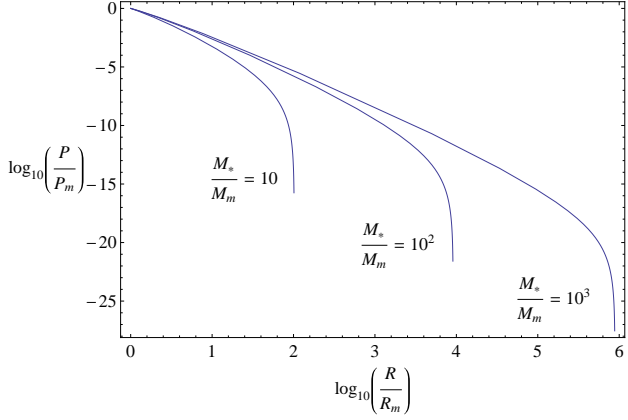}
  \caption{Same as Fig.~2, but for envelope pressure.}
  \label{fig:SMSEnvelope4}
\end{figure}

It is also apparent from Figures 2 and 3 that there is a simple approximate scaling relation between the envelope-to-core ratios of radius and mass:
\begin{equation}
\label{massradius}
{R_*\over R_m}\approx \left({M_*\over M_m}\right)^2 .
\end{equation}
As Fig.~4 shows, this result is not exact, but for mass ratios $\ga 10$ it is remarkably accurate.  We will see below that this relationship allows us to deduce a simple  evolutionary sequence for supermassive stars as they grow in mass. 

\subsection{Core--Envelope Co-evolution}

\begin{figure}
  \centering
  \includegraphics[bb=0 0 321 218,width=3.23in,height=3.23in,keepaspectratio]{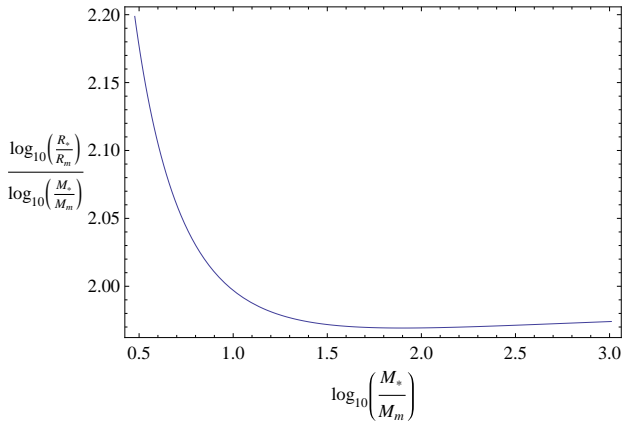}
  \caption{For envelope-to-core mass ratios in the range $10-10^3$, the ratio of envelope-to-core radius is very close to the square of the mass ratio.}
  \label{fig:SMSEnvelope2a}
\end{figure}

As in a normal star, we expect the extreme temperature-sensitivity of thermonuclear reactions to regulate the central temperature of the core, and therefore $P_c = aT_c^4/3$, to a nearly constant value as the core evolves.  Given that $x_m$ (and therefore $\eta$) is roughly constant as well, the relationship between mass and radius of the core becomes a simple scaling law:
\begin{equation}
\label{massradiuscore}
R_m = x_m \left({3\over 4 \eta}\right)^{3/4}\left({4\over \pi}\right)^{1/4} G^{1/4} P_c^{-1/4} M_m^{1/2} \propto M_m^{1/2},
\end{equation} 
which allows us to eliminate $M_m$ in favor of $R_m$ in eq.~(\ref{massradius}).  Since $M_* = \dot M_* t$ is given as a function of time, eq.~(\ref{massradius}) is readily reduced to a relation between $R_*$ and $R_m$.  A second, independent relation linking these two radii would therefore suffice to determine the co-evolution of the core and envelope.  This relationship is provided by the virial theorem.

According to the virial theorem, the total energy of a radiation-dominated star is very much smaller than the canonical value of $-GM_*^2/R_*$, $E_{\rm tot} = - E_g - E_{\rm rot}+ E_{\rm GR}$, where $E_g$ is the thermal gas energy, $E_{\rm rot}$ is the rotational kinetic energy, and $E_{\rm GR}$ is the energy correction associated with general relativistic effects (Fowler 1964).  The relativistic correction must be outweighed by the gas pressure and rotational terms in order to ensure stability, and we henceforth neglect it. If we assume that fresh matter joining the star is ``cold," so that $4 P/\rho \ll GM_*/R_*$, then accretion {\it decreases} the energy of the star at a rate $-GM_*\dot M_* / R_*$.  In addition, the star leaks energy at the Eddington limit, so that the total rate of energy loss is $-GM_*\dot M_*/ R_* - L_E(M_*) = -(GM_* \dot M_*/ R_*)(1 + R_*/R_{\rm tr})$, where we have used the definition of the trapping radius from eq.~(\ref{rtrap}).  The corrections to this energy loss due to changes in the gas or rotational energy are negligible (assuming that the rotation is well below Keplerian).  To maintain equilibrium, nuclear reactions in the core must compensate for virtually all of this energy loss.

The virial theorem also tells us the energy of the envelope and core separately, in terms of $R_m$ and $P_m$, the pressure at the matching radius.  The envelope has a negative energy equal to $- E_g(>R_m)- E_{\rm rot} (>R_m)-4\pi P_m R_m^3$ where, under our assumptions, $P_m = P_c \theta_3^4(x_m)$ is approximately constant.  [The core energy is $- E_g(<R_m)- E_{\rm rot} (<R_m)+4\pi P_m R_m^3$.] The envelope loses energy at $R_*$ and gains energy at $R_m$ by a combination of radiative diffusion, at a rate comparable to the Eddington limit for the core mass, $M_m$, and advection of energy across the core-envelope boundary.  Since $P_m R_m^3$ is of order $GM_*^2/R_*$, we can neglect the change of gas and rotational energy, and we also crudely estimate the energy per unit mass crossing the boundary to be $GM_*/R_*$ (i.e., the same as at $R_*$).  Matching the change of envelope energy to the energy loss at the surface, we obtain
\begin{displaymath}
4\pi {d\over dt}  ( P_m  R_m^3) =  \nonumber 
\end{displaymath}
\begin{equation}  
\label{envelopeenergy}
  \ \ \ \  \ \ \ \ \   {  G  M_*\dot M_*\over R_*}  \left[\left(1-{\dot M_m\over \dot M_*}\right)  + { R_*\over R_{\rm tr}}\left(1 - {M_m\over M_*} \right)\right]. \ \ \ \ \ 
\end{equation} 
If we consider $M_m/M_*$ to be a slowly varying function of time (we will see below that it scales roughly as $t^{1/3}$), we can further approximate $\dot M_m /\dot M_* \approx M_m/M_* $, which gives us the desired behavior that the left-hand side vanishes when $M_m$ approaches $M_*$, and also should be accurate when $M_m \ll M_*$.

Using equations (\ref{massradiuscore}) and (\ref{massradius}), and assuming constant $\dot M_*$, $P_c$, and $x_m$, we then obtain the approximate equation
\begin{equation}
\label{rstarODE}
{\chi\over t}  {d\over dt}{t^2\over R_*}= \left(1 - {M_m\over M_*} \right)  \left({1\over R_*} + {1\over R_{\rm tr}}\right),
\end{equation} 
where $\chi \equiv (27/4\eta^3) x_m^4 \theta_3^4(x_m)$.  We then obtain 
\begin{equation}
\label{rstar}
R_* \approx { 2\chi + {M_m \over M_*} - 1\over 1 - {M_m\over M_*}} R_{\rm tr} = { 2\chi + {M_m \over M_*} - 1\over 1 - {M_m\over M_*}} \ {\kappa \dot M_* \over 4\pi c}.   
\end{equation}   
As we guessed from our simple analysis based on radiation trapping arguments, the radius of a partially convective supermassive star is roughly proportional to the accretion rate.  Equation (\ref{rstar}), and eq.~(\ref{rstarODE}) from which it is derived, are only approximate because $\chi (x_m)$ does vary with $x_m$, which in turn varies weakly with $M_*/M_m$ (Fig.~1). But these relations show that $R_* \approx (0.4-0.8) R_{\rm tr}$ for $M_m/M_* \ga 0.1$, and is no larger than $\approx 1.2 R_{\rm tr}$ for $M_m/M_* >10^{-3}$.  For the purpose of the estimates to follow, it will suffice to take $R_*/R_{\rm tr} \approx 0.6$ for a partially convective star.

Once we have an expression for $R_*$, it is straightforward to calculate the evolution of the core.  In terms of the normalized quantities defined earlier, the core mass is given by 
\begin{equation}
\label{masscore}
M_m = 6.2\times 10^5 {x_m^{2/3}\over \eta^{1/2} (R_*/R_{\rm tr})^{2/3}} \dot m_*^{2/3} T_{c,8}^{-2/3} t_{\rm Myr}^{4/3} \ M_\odot,
\end{equation}
corresponding to the mass ratio
\begin{equation}
\label{masscore2}
{M_m \over M_*} = 0.62 {x_m^{2/3}\over \eta^{1/2} (R_*/ R_{\rm tr})^{2/3}} \dot m_*^{-1/3} T_{c,8}^{-2/3} t_{\rm Myr}^{1/3}.
\end{equation}
The core-to -envelope mass ratio increases with time, rapidly at first, then more gradually. The core radius, readily obtained from eq.~(\ref{massradius}), behaves in a qualitatively similar way.  

\section{Nuclear Burning}

\subsection{``Main Sequence" Lifetime}

Fully convective stars are thought to be well-mixed, and therefore capable of burning a large fraction of their hydrogen on the main sequence.  Assuming an H mass fraction $X=0.75$, the timescale to deplete all the hydrogen in a star of constant mass, radiating at the Eddington limit, is $2 \times 10^6$ yr.  If the star is growing at a steady rate by accretion then the lifetime is doubled, to $4\times 10^6$ yr.  For a supermassive star growing by accretion, however, there are two factors that can shorten the lifetime if the star does not become fully convective.  First, most of the mass may reside in the convectively stable envelope, which does not mix with the core, thus restricting the fuel supply.  Second, the thermonuclear luminosity required to maintain equilibrium exceeds the Eddington limit, because it must neutralize the binding energy of the freshly accreted matter in addition to replacing the lost radiation.  The required luminosity is 
\begin{equation}
\label{Lnuc}
L_{\rm nuc} \approx  \left( 1 + {R_{\rm tr}\over R_*} \right) L_E(M_*) \approx {8\over 3} \ {4\pi G\dot M c \over \kappa} t
\end{equation}
for a partially convective star with $M_m/M_* \ga 0.1$, and $\approx L_E$ for a fully convective star.  To obtain the total nuclear energy required over the lifetime of the star, we integrate $L_{\rm nuc}$ over time.   Comparing the total energy output with the fuel supply contained in the core, we obtain an estimate of the nuclear burning timescale,   
\begin{equation}
\label{tnuc}
t_{\rm nuc} \approx 1.5 \times 10^6 {M_m\over M_*} \ {\rm yr}. 
\end{equation}
 
We can now estimate the conditions inside a partially convective star at the end of hydrogen burning, and determine whether the star becomes fully convective.  Substituting eq.~(\ref{tnuc}) into eq.~(\ref{masscore2}), with $x_m^{2/3}/\eta^{1/2} \approx 0.9 $ and $R_*/R_{\rm tr} \approx 0.6$, we find
\begin{equation}
\label{masscore3}
{M_m \over M_*} \approx 0.9  \dot m_*^{-1/2} T_{c,8}^{-1} .
\end{equation}
The core mass is independent of the accretion rate,
\begin{equation}
\label{masscore4}
M_m  \approx 1.2 \times 10^6   T_{c,8}^{-2} \ M_\odot,
\end{equation}
while the total mass is
\begin{equation}
\label{masscore4}
M_*  \approx 1.4 \times 10^6   \dot m_*^{1/2} T_{c,8}^{-1} \ M_\odot,
\end{equation}
According to these estimates, a supermassive star will burn up its core hydrogen before reaching a fully convective state if $\dot m_* > 0.7  T_{c,8}^{-2}$. 
 
\subsection{Central Temperature}

We have shown that the thermal relaxation of a growing supermassive star is sensitive to the central temperature of the convective core, which we have assumed to be constant in anticipation of the usual nuclear thermostatic effect. Up to now we have retained this temperature as a parameter: let us now estimate it for the dominant reactions likely to be occurring in these systems.

The only thermonuclear energy source that can support a supermassive star is the CNO cycle (Hoyle \& Fowler 1963a).  At temperatures of interest the cycle reaches equilibrium quickly, hence we use the equilibrium reaction rate (Clayton 1983), assuming a hydrogen mass fraction $X=0.75$ but leaving the CNO mass fraction $Z_{\rm CNO}$ as a parameter.  We have also considered the possible role of the ``hot" CNO mode (Mathews \& Dietrich 1984) and the high-temperature saturation of the energy generation rate (e.g., Narayan \& Heyl 2003), but have found both of these effects to be unimportant at temperatures present in the core, due to the very low central densities of supermassive stars ($\rho_c = 0.04 \eta^{3/4} m_{m,6}^{-1/2} T_{c,8}^3$ g cm$^{-3}$).    

The energy generation rate within the convective core can be written in the form
\begin{displaymath}
 L_{\rm c} =   \nonumber
\end{displaymath}
\begin{equation}
\label{Lcore}
 \ \ \ \\  \ A {m_{m,6}^{1/2} T_{c,8}^{7/3} \over \eta^{3/4}}\nonumber  
 \int_0^{x_m} x^2 \theta_3^{16/3}(x) \exp \left[-{B\over  T_{c,8}^{1/3}\theta_3^{1/3}(x)}\right] dx,
\end{equation}
where $A$ and $B$ are constants.  For a fully convective star, use $\eta = 4.8$, replace $m_{m,6}$ by $m_{*,6}$, and set $x_m = 6.89685$.  In addition to computing the nuclear energy generation in the core, we will need to check whether any significant energy generation occurs in the envelope.

To determine the equilibrium central temperature, we set $L_c$ equal to the required energy generation rate given by eq.~(\ref{Lnuc}).  We use $A= 1.10 \times 10^{65} Z_{\rm CNO}$ erg s$^{-1}$, $B = 32.81$ in eq.~(\ref{Lcore}) to obtain $L_{c, {\rm CNO}}$.  For all relevant values of $M_m/M_*$, we can verify that essentially all nuclear burning is confined to the core.  

Figure 5 shows the equilibrium value of $T_{c,8}$ as a function of $\log_{10}(Z_{\rm CNO}/ m_{*,6}^{1/2})$. For $m_{*, 6} \sim O(1)$ and $Z_{\rm CNO}$ close to solar abundances ($\sim 0.01$), we reproduce the result from Hoyle \& Fowler (1963a) that $T_{c,8} \sim 0.7$.  Low temperatures favor more thermally relaxed stars (larger $M_m/M_*$) for a given accretion rate and elapsed time, primarily because the core radius is inversely proportional to $T_c$. On the other, hand, for low CNO abundances ($\la 10^{-4}$) that may be present at high redshifts where supermassive stars might be forming, $T_c$ can exceed $10^8$ K, possibly exceeding $2\times 10^8$ under extreme conditions of low metallicity.  Such high temperatures would imply that supermassive stars are never thermally relaxed, even for modest accretion rates $\la 1\  M_\odot$ yr$^{-1}$. 

\begin{figure}
  \centering
  \includegraphics[bb=0 0 240 262,width=3.23in,height=3.23in,keepaspectratio]{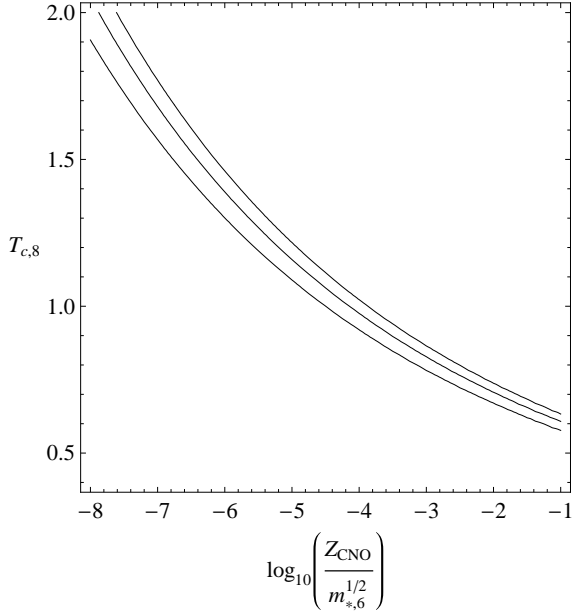}
  \caption{Central temperature of the convective core in units of $10^8$ K, as a function of the CNO mass fraction $Z_{\rm CNO}$ divided by $m_{*,6}^{1/2}$.  Top, middle, and bottom curves are for envelope-to-core mass ratios $M_*/M_m= 3,\ 10^2$, and $10$, respectively.  (Note that $T_{c,8}$ is not a monotonic function of $M_*/M_m$.) }
  \label{fig:SMSEnvelope6}
\end{figure}

If the star is totally devoid of CNO nuclei, hydrogen burns primarily via the PP III branch of the proton-proton chain (Clayton 1983).  Accounting for neutrino losses, we have $A= 3.71 \times 10^{43}$ erg s$^{-1}$, $B = 7.28$ in eq.~(\ref{Lcore}).  It is straightforward to show that the proton-proton chain can never meet the energy requirements of a supermassive star, nor can helium burning via the triple-$\alpha$ process.  

One can debate the likelihood that a supermassive star with such low CNO  abundance ($Z_{\rm CNO} \la 10^{-8}$) could ever form under cosmological conditions (e.g., Trenti \& Stiavelli 2009; Trenti, Stiavelli \& Shull 2009).  The point may be moot, however, because a supermassive star will create its own metals.  To see this, we estimate the production rate of CNO elements via the triple-$\alpha$ process in the core:
\begin{displaymath}
\dot M_{\rm CNO} =  
\end{displaymath}
\begin{equation}
\label{MdotCNO}
\ \ 3 \times 10^3 T_{c,8}^3 \int_0^{x_m} x^2 \theta_3^6 (x) \exp \left[-{44.03\over  T_{c,8}\theta_3(x)}\right] dx \ M_\odot \ {\rm yr}^{-1} .
\end{equation} 
At $T_{c,8} \sim 2$, the core produces $\sim 10^{-7} M_\odot$ yr$^{-1}$, and at $T_{c,8} \sim 3$ the CNO production rate is increased $> 10^{-3} M_\odot$ yr$^{-1}$.  If the star can persist for a few thousand years in Kelvin contraction at $T_{c,8} \la 3$, it should be able to produce enough CNO elements to support itself by burning hydrogen.

\section{Black Hole Formation and Growth}

\subsection{Formation of a Seed Black Hole}

After exhausting its hydrogen, the core of a supermassive star will contract and heat up until it suffers catastrophic neutrino losses and collapses. If the mass shell $M$ encloses a specific angular momentum less than $\sim GM/c$, it can collapse directly to a black hole without any angular momentum transport.  Approximating the convective core as a uniform sphere of density $\rho_c$ and radius $R_m$, we can estimate the maximum ratio of angular velocity, $\Omega$, to local Keplerian angular velocity, $\Omega_K = (GM/R^3)^{1/2}$, consistent with direct black hole formation:
\begin{equation}
\label{omegak}
\varepsilon(R) \equiv \left({\Omega\over\Omega_K}\right)^2 < {GM(R)\over c^2 R} \sim 0.01 m_{m,6}^{1/2} T_{c,8} \left({R\over R_m}\right)^2 .
\end{equation}
Note that this constraint becomes more stringent at smaller radii.  If, for example, the convection drives the core to solid body rotation, $\Omega \approx$ const., then $\varepsilon (R)$ is also constant and the core might satisfy condition (\ref{omegak}) at large radii but not near the center.  Given that the supermassive star is formed from gas that probably has substantial rotation, this stringent constraint suggests that the formation of the black hole depends on angular momentum transport in the collapsing gas.  Since the collapsing gas within any mass shell is strongly self-gravitating, angular momentum transport can be driven efficiently by global gravitational torques and turbulence resulting from nonaxisymmetric gravitational instabilities, as discussed in Begelman et al.~(2006) --- although magnetic torques could also be important.  

The transport of angular momentum outward is unavoidably accompanied by the outward transport of energy (Blandford \& Begelman 1999).  An unknown fraction of this liberated energy might escape the system via a jet punching through the star (as in the collapsar model of gamma-ray bursts). Neutrino losses might also be important, particularly when the black hole mass is small.  The remainder of the energy must pass through the star on its way out, presumably driving a strong convective flux through the outer core and surrounding envelope (Blandford \& Begelman 2004).  In associating the convective luminosity with the rate of black hole growth, $L_{\rm BH} = \epsilon M_{\rm BH} c^2$, we fold in these possible inefficiencies through the factor $\epsilon$.  We recognize the standard efficiency of black hole accretion by using the normalization $\epsilon = 0.1 \epsilon_{-1}$, with the understanding that $\epsilon_{-1}$ might be $\ll 1$.
 
Initially, the liberated energy flux is trapped in the star, causing it to inflate.  As we show below, this happens so quickly that $M_*$ can be taken to be constant during this process.  Since $R \propto M^{1/2} T_c^{-1}$ for a radiation-dominated, convective body, the core temperature drops rapidly as the star expands, as does the total binding energy.  An upper limit to the energy that can be absorbed without dispersing the star is the initial binding energy, $E_b = E_g + E_{\rm rot}$ (neglecting general relativistic corrections).  For a fully convective star, $E_g = 2.3 \times 10^{55} m_{*,6} T_{c,8}$ erg while $E_{\rm rot} = 3.4 \times 10^{57}m_{*,6}^{3/2} T_{c,8} \langle \varepsilon \rangle_W$ erg, where $\langle \varepsilon \rangle_W$ is the ``potential energy-weighted" mean rotation parameter $W^{-1} \int \varepsilon GM(R)dM/R $.  This leads to a maximum ``seed" black hole mass that can grow during the intial expansion phase:
\begin{equation}
\label{seedmass}
M_{\rm seed} = {E_b\over \epsilon c^2} = 130 { m_{*,6} T_{c,8}\over \epsilon_{-1}}\left(1 + 150 m_{*,6}^{1/2} \langle \varepsilon \rangle_W \right) \ M_\odot.
\end{equation}   
The estimate of $E_b$ (and thus $M_{\rm seed}$) is easily generalized to the case where the supermassive star is not fully convective when the black hole starts to grow.

To estimate the timescale for this initial growth, we apply the modified Bondi accretion model described in Begalman et al.~(2006, 2008).  Because the accretion radius of the seed black hole is small compared to the radius of the star, we can use the central pressure and density as approximate outer boundary conditions for the flow.  The standard Bondi (1952) formula for the accretion rate is reduced by a factor  $\sim \epsilon^{-1} (c_s/c)^2$, where $c_s = (4 P_c/ 3\rho_c)^{1/2} $ is the adiabatic sound speed outside the accretion radius.  This factor is necessary to ensure that the convective flux escaping from the accretion flow does not exceed the saturated value $\sim P_c c_c$ (Gruzinov 1998; Blandford \& Begelman 1999; Narayan, Igumenshchev \& Abramowicz 2000; Quataert \& Gruzinov 2000). For a black hole mass of $m_{\rm BH} M_\odot $, the growth time is 
\begin{equation}
\label{mdotbh}
t_{\rm BH} = {M_{\rm BH}\over \dot M_{\rm BH}} =  450\  \epsilon_{-1}  m_{\rm BH}^{-1}  m_{*,6}^{3/4} T_{c,8}^{-5/2}  \ {\rm yr}, 
\end{equation}   
implying that a black hole of modest initial mass will quickly grow to $M_{\rm seed}$.

The growth of the black hole cannot completely disrupt the star, because this would interrupt the fuel supply and stop accretion.  Once the black hole reaches a mass $M_{\rm seed}$, the convective envelope expands at approximately constant mass, in response to the black hole power
\begin{equation}
\label{Lbhseed}
L_{\rm BH} = 2 \times 10^{42} \ \epsilon_{-1}  m_{\rm seed}^2  m_{*,6}^{-3/4} T_{c,8}^{5/2} \ {\rm erg \ s}^{-1} \propto R_*^{-5/2}. 
\end{equation}  
The star continues to expand as long as $L_{\rm BH}$ exceeds the Eddington limit for the star, $L_E = 1.4 \times 10^{44} M_{*, 6}$ erg s$^{-1}$. Once the limit is reach, the black hole--star system should come into a kind of equilibrium that we have dubbed a ``quasistar" --- a bloated convective envelope powered by black hole accretion at the center.

\subsection{Growth Inside Quasistar}

As discussed in previous publications (Begelman et al.~2006, 2008), under suitable initial conditions a black hole can grow rapidly and by a large factor inside a quasistar.  The growth occurs at a highly super-Eddington rate (i.e., with mass e-folding in much less than the Salpeter time, $\sim 40$ Myr) because the accretion power of the black hole is regulated at the Eddington limit corresponding to $M_*$, which is much larger than $M_{\rm BH}$.  The main requirement for stable growth is that the ratio of black-hole to envelope mass be smaller than $\sim 0.01$ (with some dependence on envelope abundances, via the opacity); otherwise the effective temperature becomes so low that the envelope cannot radiate away the flux, and it disperses.  We refer readers to Begelman et al.~(2008) for detailed discussion of quasistar structure and evolution.

To calculate the co-evolution of the black hole and quasistar, we apply the model described in section 4 of Begelman et al.~(2008), with the two feedback efficiency parameters, $\alpha$ and $\epsilon$, set equal for simplicity.  Integrating eq.~(51) of Begelman et al.~(2008) and using the quasistar mass as the independent variable rather than time, we obtain
\begin{equation}
\label{Mbhqs}
m_{\rm BH} =  m_{\rm seed} + {1.3 \times 10^4 \over \epsilon_{-1} \dot m_*} \left( m_{*,6}^2 - m_{i,6}^2 \right) , 
\end{equation}  
where $m_{i,6} \sim \min \left[4 \dot m_*, 1.4 \dot m_*^{1/2}T_{c,8}^{-1}  \right]$
is the initial mass of the quasistar in units of $10^6 M_\odot$, from eq.~(\ref{masscore4}).

From eq.~(11) of Begelman et al.~(2008) (with parameters set to their fiducial values), the effective temperature of the quasistar's photosphere (in units of $10^3$ K) is given by
\begin{equation} 
\label{Tphot1}
T_{\rm ph, 3} = 2 \times 10^2  \epsilon_{-1}^{-1/5} m_{\rm BH}^{-2/5} m_{*,6}^{7/20} ,
\end{equation}
which decreases with time as the black hole and quasistar grow (Begelman et al.~2008).  The quasistar disperses and black hole growth stops when $T_{\rm ph}$ reaches $T_{\rm min}\approx 4000$ K.  In contrast to a contracting protostar or red giant envelope which manages to maintain dynamical equilibrium while evolving at close to the minimum effective temperature (the ``Hayashi track"), the quasistar loses equilibrium and disperses at this point because of the nature of the black hole energy source as well as the fact that it is supported by radiation pressure rather than gas pressure.  This behavior is discussed at length in Begelman et al.~(2008).

Given the estimates above for $M_{\rm seed}$ and $M_i$, it is clear that the black hole is generally able grow by a large factor before the quasistar disperses.  Assuming that the final black hole mass, $m_{f, {\rm BH}}$ is much larger than $m_{\rm seed}$, we find 
\begin{equation}
\label{Mbhfqs}
m_{f,{\rm BH}} =  {1.3 \times 10^4 \over \epsilon_{-1}^{1/2}} \left(  {4600 \ {\rm K}\over T_{\rm min}} \right)^{5/2} m_{f,6}^{7/8} \approx   {2\times 10^4 \over \epsilon_{-1}^{1/2}}  m_{f,6}^{7/8}, 
\end{equation}  
where $M_f$ is the final mass of the quasistar.  For modest accretion rates $\dot m_* \la O(1)$, the quasistar does not grow much between its formation and dispersal, $M_f \approx M_i$, and for $\dot m_* \sim O(1)$ the black hole left behind when the quasistar disperses has a mass in the range $\sim (10^4-10^5) \epsilon_{-1}^{-1/2} M_\odot$. But for $\dot m_* \gg 3 \epsilon_{-1}^{8/9} T_{c,8}^{14/9}$, the final quasistar mass is considerably larger, $m_{f,6} \sim 1.4 \epsilon_{-1}^{4/9} \dot m_*^{8/9}$ and the final black hole mass is $m_{f, {\rm BH}} \sim 3 \times 10^4 \epsilon_{-1}^{-1/9} \dot m_*^{7/9}$. 

\section{Summary and Discussion}

We have revisited the theory of supermassive stars, taking into account the fact that their short nuclear burning timescales ($\leq 4$ Myr) imply that the matter forming them must accumulate rapidly.  If the accumulation of matter is rapid enough (typically, more than a few solar masses per year), then the star never has time to become thermally relaxed and fully convective. Instead, its structure consists of a convective core containing a minority of the mass and occupying a small fraction of the stellar volume, surrounded by a convectively stable envelope with the specific entropy increasing outward as enclosed mass to the 2/3 power. We have calculated the structures of these envelopes --- for which we propose the term ``hylotropes" (following a suggestion by A.~Accardi and G.~Lodato) --- and the conditions for matching them to the convective cores, and have thus devised a theory for the evolution of partially convective supermassive stars.  

Partially convective supermassive stars are strikingly different from their fully convective counterparts (described by $n=3$ polytropes) in a number of respects.  They do not follow the same mass-radius relation as fully convective stars.  Instead of the scaling $R \propto M^{1/2}$, which applies for a fully convective supermassive star with a central temperature regulated by thermonuclear reactions, our partially convective models have a radius proportional to the (assumed constant) accretion rate.  Physically, the stellar radius is comparable to the ``trapping radius" within which radiation is unable to escape inward advection by the accumulating gas (Begelman 1978, 1979). Also, their nuclear-burning (``main sequence") lifetimes can be shorter because the fuel in the convectively stable envelope is never mixed into the core. 

Hydrogen-burning supermassive stars in the mass range considered here ($\ga 10^6 M_\odot$) can exist only if stabilized by rotation or some other ``stiff" form of energy (such as magnetic fields or turbulence) (Fowler 1964, 1966).  The minimal level of rotation required to cancel the general-relativistic instability is not dynamically significant.  However, the angular momentum may well be high enough to inhibit the formation of the black hole, once the core runs out of hydrogen and collapses.  We have argued that the initial mass of the seed black hole, if limited by the energy released in forming it, may be as small as a few hundred solar masses --- i.e., only $10^{-4}$ of the total stellar mass.  However, there is likely to be a second phase of rapid black hole growth once the inflated stellar envelope reaches equilibrium as a red giant-like quasistar.  We therefore estimate that supermassive stars should leave behind black holes with typical masses of a few percent that of the star, i.e., roughly $\sim 10^4 - 10^5 M_\odot$.  

These estimates are admittedly quite uncertain, mainly because we do not understand how energy liberated by black hole accretion couples to the remainder of the star.  We cannot rule out the possibility that most of the energy punches through the star in a pair of jets (in a fashion similar to the collapsar model  of gamma-ray bursts [Woosley 1993]), in which case both the seed and final black hole masses could be much larger.   We note that while our estimate of the initial black hole mass is qualitatively similar to values (few tens of $M_\odot$) predicted by Begelman et al.~(2006), our new value for the predicted envelope mass at the time of black hole formation is 2--3 orders of magnitude larger.  The reason for this discrepancy is that Begelman et al.~(2006) failed to account for the rapid growth of the convective core due to heating by efficient thermonuclear reactions in the CNO cycle.  (Effectively, we considered pure Pop III abundances, for which thermonuclear energy release is too weak to stall the formation of the seed hole.)   

Supermassive stars can be strong sources of far-UV radiation, provided that their spectra are not severely degraded by reprocessing at radii beyond $R_{\rm tr}$.  Taking the photospheric radius to be $0.6 R_{\rm tr}$, we find an effective temperature $T_{\rm eff} = 1.2 \times 10^5 m_{*,6}^{1/4} \dot m_*^{-1/2}$ K for a partially convective star of mass $M_*$, corresponding to $1.3 \times 10^5 \dot m_*^{-3/8} T_{c,8}^{-1/4}$ K at the end of hydrogen-burning.  However, the infalling gas joining a supermassive star is already very optically thick at $R_{\rm tr}$ ($\tau \sim c/v$, where $v$ is the infall speed), implying that most of the radiation escapes from further out and is likely to be softer. On the other hand, the opacity is strongly scattering-dominated, so the color temperature could be several times higher than the effective temperature.  The hardest radiation is likely to escape from the polar regions, where rotational effects decrease the density. 

Assuming a blackbody spectrum with the effective temperature estimated above and $\dot m_* = T_{c,8} = 1$, we find that most of the radiation is capable of ionizing hydrogen ($E > 13.6$ eV), with an ionizing luminosity $L_{\rm ion}\approx 2\times 10^{44}$ erg s$^{-1}$ and photon flux ${\cal N}_{\rm ion} \approx 4 \times 10^{54}$ s$^{-1}$. The star radiates into the Lyman--Werner band ($11.2 < E < 13.6$ eV), capable of dissociating H$_2$, with $L_{\rm diss}\approx 5\times 10^{42}$ erg s$^{-1}$ and ${\cal N}_{\rm diss} \approx 3 \times 10^{53}$ s$^{-1}$.  To produce an interesting mean intensity of dissociating radiation, $J_\nu > 10^{-22}$ erg cm$^{-2}$ s$^{-1}$ Hz$^{-1}$ sr$^{-1}$ between $z\sim 12$ and $z\sim 6$, would require (very roughly) a comoving density of supermassive stars of at least $\sim 10^{-4}$  Mpc$^{-3}$.  Given the short lifetimes of these objects, it is not clear how plausible this is.  We will investigate cosmological scenarios for the formation of supermassive stars in a future publication.

\section*{Acknowledgments}
This paper grew out of a conversation with Lars Bildsten during the 2007 program ``Star Formation Through Cosmic Time" at the Kavli Institute for Theoretical Physics, University of California at Santa Barbara.  The research was supported in part by the National Science Foundation under Grants PHY~0551164 (KITP) and AST~0307502, and by NASA's Astrophysics Theory and Beyond Einstein Foundation Science programs under grants NNG04GL01G and NNG05GI92G.  I also thank Phil Armitage and Elena Rossi for numerous discussions and Alice Accardi and Giuseppe Lodato for suggesting the term ``hylotrope" and providing its etymology.


\begin{thebibliography}{}


\bibitem[Begelman(1978)]{begelman78} Begelman M.~C., 1978, MNRAS, 184, 53

\bibitem[Begelman(1979)]{begelman79} Begelman M.~C., 1979, MNRAS, 187, 237
 
\bibitem[Begelman, Rossi \& Armitage(2008)]{Be08} Begelman M.~C., Rossi E., Armitage P.~J., 2008, MNRAS, 387, 1649

\bibitem[]{} Begelman M.~C., Shlosman I., 2009, ApJ, in press (arXiv:0904.4247)
 
\bibitem[Begelman, Volonteri \& Rees(2006)]{Be06} Begelman M.~C., Volonteri M., Rees M.~J., 2006, MNRAS, 370, 289 

\bibitem[]{}Bisnovatyi-Kogan G.~S., Zel'dovich Ya.~B., Novikov I.~D., 1967, Sov.~Astron., 11, 419
Publication Date:	
	12/1967

\bibitem[Blandford \& Begelman(1999)]{bla99} Blandford R.~D., Begelman M.~C., 1999, MNRAS, 303, L1

\bibitem[Blandford \& Begelman(2004)]{bla04} Blandford R.~D., Begelman M.~C., 2004, MNRAS, 349, 68

\bibitem[Bondi(1952)]{bo52} Bondi H., 1952, MNRAS, 112, 195

\bibitem[Bromm \& Loeb 2003]{bromm03} Bromm V., Loeb A., 2003, ApJ, 596, 34

\bibitem[Burbidge(1958)]{bur58} Burbidge G.~R., 1958, in Paris Symposium on Radio Astronomy (IAU-URSI), 541

\bibitem[Clayton (1983)]{cla83} Clayton D.~D. 1983, Principles of Stellar Evolution and Nucleosynthesis (Chicago: University of Chicago Press)
 
\bibitem[Fowler(1964)]{Fo64} Fowler W.~A., 1964, RMP, 36, 545

\bibitem[Fowler(1966)]{Fo66} Fowler W.~A., 1966, ApJ, 144, 180 

\bibitem[Fuller et al.(1986)]{Fu86} Fuller G.~M., Woosley S.~E., Weaver T.~A., 1986, ApJ, 307, 675

\bibitem[Gruzinov(1998)]{Gr98} Gruzinov A., 1998, unpublished manuscript, astro-ph/9809265 

\bibitem[Hoyle \& Fowler(1963a)]{HoFo63a} Hoyle F., Fowler W.~A., 1963a, MNRAS, 125, 169 

\bibitem[Hoyle \& Fowler(1963b)]{HoFo63b} Hoyle F., Fowler W.~A., 1963b, Nature, 197, 533 

\bibitem[Iben(1963)]{Ib63} Iben I., 1963, ApJ, 138, 1090

\bibitem[]{}Levine R., Gnedin N.~Y., Hamilton A.~J.~S., Kravtsov A.~V., 2008, ApJ, 678, 154
 
\bibitem[Mathews \& Dietrich (1984)]{Ma84} Mathews G.~J., Dietrich, F.~S. 1984, ApJ, 287, 969 

\bibitem[Narayan \& Heyl (2003)]{Na03} Narayan, R., Heyl J.~S. 1983, ApJ, 599, 419 

\bibitem[Narayan, Igumenshchev \& Abramowicz(2000)]{Na00} Narayan R., Igumenshchev I.~V., Abramowicz M.~A., 2000, ApJ, 539, 798

\bibitem[Ozernoy \& Usov(1971)]{OzUs71} Ozernoy L.~M., Usov V.~V., 1971, Ap\&SS, 13, 3 

\bibitem[Quataert \& Gruzinov(2000)]{Qu00} Quataert E., Gruzinov A., 2000, ApJ, 539, 809

\bibitem[]{}Regan J.~A., Haehnelt M.~G., 2009, MNRAS, 396, 343 

\bibitem[]{} Trenti M., Stiavelli M., 2009, ApJ, 694, 879

\bibitem[]{} Trenti M., Stiavelli M., Shull J.~M. 2009, ApJ, 700, 1672
 
\bibitem[]{}Wise J.~H., Turk M.~J., Abel T., 2008, ApJ, 682, 745

\bibitem[]{} Woosley S.~E., 1993, ApJ, 405, 273 

\end{thebibliography}
\end{document}